\documentclass[aps,prl,twocolumn,floatfix]{revtex4}
\usepackage{epsfig}

\begin{document}

\title{Towards New Tests of Strong-field Gravity with Measurements of Surface Atomic Line Redshifts from Neutron Stars}
\author{Simon DeDeo}
\affiliation{Department of Astrophysical Sciences, Princeton University, Princeton, NJ 08544}
\author{Dimitrios Psaltis}
\affiliation{School of Natural Sciences, Institute for Advanced Study, Einstein Drive, Princeton, NJ 08540}

\begin{abstract}
In contrast to gravity in the weak-field regime, which has been
subject to numerous experimental tests, gravity in the strong-field
regime is largely unconstrained by observations. We show that gravity
theories that cannot be rejected by solar system tests but that
diverge from general relativity in the strong-field regime predict
neutron stars with significantly different properties than their
general relativistic counterparts. In particular, the range of
redshifts of surface atomic lines predicted by such gravity theories
is significantly larger than the uncertainty introduced by our lack of
knowledge of the equation of state of ultra-dense matter. Measurements
of such redshifted lines with current X-ray observatories such as
Chandra and XMM-Newton can thus provide interesting new constraints on
strong-field gravity.
\end{abstract}

\maketitle

The rapid variability and high luminosities of accreting compact
objects strongly suggest that their high-energy emission originates
very deep in their gravitational potentials. As a result, the
properties of their X-ray and $\gamma$-ray emission can, in principle,
be used to test directly the strong-field regime of a gravity
theory. Unfortunately, there are many astrophysical considerations
that complicate the modeling of the emission of accreting neutron
stars and black holes and it is difficult to disentangle the effects
of gravity from those due to astrophysical processes. For this reason
tests of gravity involving high-energy phenomena around compact
objects have received little attention in the past. This is despite
the fact that the behavior of gravity in the strong-field regime is
largely unconstrained by observations.

The properties of a neutron star and its external spacetime, in
contrast to the high-energy phenomena above its surface, are
determined solely by the strong-field behavior of gravity and the
equation of state. To date, studies of neutron stars have focused on
better determining the equation of state, whose details are still
debated. We will argue, however, that the uncertainty in the behavior
of strong-field gravity introduces much greater differences in
neutron-star properties than do current uncertainties in the
neutron-star equation of state. Furthermore, new observational
developments allow for the measurement of the gravitational redshift
of surface atomic lines, which is the cleanest source of information
about the mass, radius, and spin of neutron stars. The first such
observations, with Chandra and XMM-Newton, have already been
reported~\cite{cpm02,spzt02}.

Testing rigorously General Relativity (GR) in the strong-field regime
requires a general framework of which GR is a subset; this would be
equivalent to the Parametrized Post-Newtonian formulation in the
weak-field. Such a framework is yet to be constructed. Thus, in order
to provide a proof of principle for the efficacy of the proposed
tests, we will investigate here the constraints that can be imposed on
a parametrized subclass of scalar-tensor theories (see, e.g.,
Ref. ~\cite{de92}). As an additional example, we will also consider
the limits that can be placed on Rosen's bimetric theory \cite{r73};
although binary-pulsar experiments have already excluded the bimetric
theory \cite{we77}, our results demonstrate the potential of the
proposed tests for constraining gravity theories.

We consider first \emph{scalar-tensor theories}. The general class of
scalar-tensor theories, where the gravitational force felt by matter
is mediated by both a rank-two tensor, $g_{\mu\nu}$, and a scalar
field, $\phi$, is one of the most natural extensions of Einstein's
theory. The action is \cite{de92}
\begin{displaymath}
S = \frac{1}{16\pi G_{*}}\int d^4x \sqrt{-g_{*}}(R_{*}-2g^{\mu\nu}_{*}\phi_{,\mu}\phi_{,\nu})
\end{displaymath}\begin{equation}
\label{lagrangian}
+ S_m[\Psi_m;A^2(\phi)g_{*\mu\nu}].
\end{equation}
Here $\Psi_m$ refers collectively to all matter fields other than
$\phi$, $G_*$ is a dimensional constant, and $A(\phi)$ is a function
of $\phi$. The tensor $g_{*\mu\nu}$ is the metric in the Einstein
frame and the Brans-Dicke frame tensor is
$g_{\mu\nu}=A^2(\phi)g_{*\mu\nu}$. For $A(\phi)$ unity, we recover GR.

Solar system experiments constrain the coupling function $A$ to be
very flat at the cosmological value of the scalar field $\phi$. If,
however, the scalar field inside a compact object fluctuates far
enough away from its cosmological value to discover a steep part of
$A$, the system will be able to reach the more energetically favorable
configuration with large non-zero $\phi$ near the center of the
object. This non-perturbative effect leads to neutron stars that are
significantly more massive and larger than their GR counterparts, but
it is invisible to solar system experiments that probe the weak-field
regime. This is the phenomenon of ``spontaneous scalarization''
discovered by Damour \& Esposito-Far\`{e}se \cite{de93}, who used the
Hulse-Taylor pulsar to put constraints on the strong-field regime of
such theories.

Following Damour \& Esposito-Far\`{e}se \cite{de93}, we choose
\begin{equation}
A(\phi)=e^{\frac{1}{2}\beta\phi^2},
\end{equation}
where $\beta$ is a real number. 
Spontaneous scalarization occurs only when
$\beta\leq-4.85$. Throughout this paper we will be interested only in
solutions where $\phi_0$, the cosmological value of the scalar field,
is zero and $A(\phi_0)=1$.

Following Ref. \cite{de93}, we write the relativistic equations of
stellar structure for a {\bf non-rotating} neutron star in
scalar-tensor gravity. We write the metric as
\begin{eqnarray}
ds_*^2 &=& -e^{\nu(r)}dt^2+\left[1-\frac{2\mu(r)}{r}\right]^{-1}dr^2\nonumber\\
& &+ r^2(d\theta^2+\sin^2\theta d\phi^2),
\label{metric}
\end{eqnarray}
and describe the matter fields as a perfect fluid in the physical frame, i.e.,  
\begin{equation}
T_{\mu\nu}=(\rho+p)u_{\mu}u_{\nu}+pg_{\mu\nu}.
\end{equation}
Here $p$ is the pressure, $\rho$ is the density, and $u_{\mu}$ is the
four-velocity of the fluid.
The differential equations to be integrated are then~\cite{h98}
\begin{equation}
\label{eqs}
\mu^{\prime} = 4\pi G_*r^2A^4\rho+\frac{1}{2}r(r-2\mu)\psi^2,
\end{equation}\begin{equation}
\nu^{\prime} = 8\pi g_*\frac{r^2A^4p}{r-2\mu}+r\psi^2+\frac{2\mu}{r(r-2\mu)},
\end{equation}\begin{equation}\phi^{\prime} = \psi,\end{equation}
\begin{eqnarray}
\psi^{\prime} &=& 4\pi G_* \frac{rA^4}{r-2\mu}[\alpha(\rho-3p)+r(\rho-p)\psi]
\nonumber\\
&&+ \frac{\mu}{r(r-2\mu)}\psi,
\label{harry}
\end{eqnarray}
\begin{equation}N^{\prime} = 4\pi nA^3r^2\left(1-\frac{2\mu}{r}\right)^{-1/2},
\end{equation}
\begin{displaymath}p^{\prime} = -(\rho+p)\left[4\pi G_*\frac{r^2A^4p}{r-2\mu}+\frac{1}{2}r\psi^2\right]
\end{displaymath}
\begin{equation}
\label{eqs2}
-(\rho+p)\left[\frac{\mu}{r(r-2\mu)}+\alpha\psi\right].
\end{equation}
Here $N$ is the baryon number, $n$ is the number density, and primes
denote derivatives with respect to $r$. We supplement these equations
with a (zero temperature) equation of state, $p=p(\rho)$ and
$n=n(\rho)$.

We choose two commonly used equations of state, which cover a broad
subset of the wide range discussed in Cook et al. \cite{cst94}.  In
order of increasing stiffness, these are EOS~A
\cite{eosa} and EOS~UU \cite{eosuu}. Neutron star models computed with
these two equations of state bracket the uncertainty introduced by our
inability to calculate from first principles the properties of
ultra-dense matter, when condensates or unconfined u-d-s quark matter
is not taken into account.

Having chosen an equation of state, we then integrate the system of
equations (\ref{eqs})--(\ref{eqs2}) from the center of the star, where
we specify $\mu(0)=\nu(0)=N(0)=0,\phi(0)=\phi_c,$ and
$\rho(0)=\rho_c$, to the surface, where $p(R)=0$. We then integrate
equations (\ref{eqs})--(\ref{harry}) from the surface of the star to
infinity to determine the form of the metric exterior to the star.
We note that the Arnowitt-Deser-Misner (ADM) mass felt by an observer
far away is in general different to the mass that contributes to the
redshift.

In addition to considering the scalar-tensor theory, we also
investigate the bimetric theory, proposed by Rosen \cite{r73}. This
involves, in addition to the dynamical metric $g_{\mu\nu}$, a
nondynamical flat metric $\eta_{\mu\nu}$.

The metric $g_{\mu\nu}$ in Rosen's theory may be written as
\begin{displaymath}
ds^2 = -e^{-2m_{\Phi}(r)/r} dt^2
\end{displaymath}\begin{equation}
+ e^{2m_{\Lambda}(r)/r} [dr^2+r^2(d\theta^2+\sin^2 \theta d\phi^2)],
\end{equation}where $m_{\Phi}(r)$ and $m_{\Lambda}(r)$ are given, interior to the star, by
\begin{equation}
m_{\Phi}(r) = 4\pi \int^r_0 e^{\Phi+3\Lambda} (\rho+3p)r^2 dr,
\end{equation}
\begin{equation}
m_{\Lambda}(r) = 4\pi \int^r_0 e^{\Phi+\Lambda} (\rho-p)r^2 dr,
\end{equation}and $\Phi$ and $\Lambda$ are given by the field equations
\begin{equation}
\Phi^{\prime} = Gm_{\Phi}(r)/r^2,
\end{equation}
\begin{equation}
\Lambda^{\prime} = -Gm_{\Lambda}(r)/r^2.
\end{equation}
The equation of hydrostatic equilibrium is
\begin{equation}
p^{\prime} = -(\rho+p)\Phi^{\prime}.
\end{equation}
Outside the star, $m_{\Phi}(r)=m_{\Phi}(R)\equiv M_{\Phi}$ and
$m_{\Lambda}(r)=m_{\Lambda}(R)\equiv M_{\Lambda}$ are
constants. $M_{\Phi}$ is equal to the Kepler-measured mass at large
distances. We specify an equation of state, $p=p(\rho)$, and the boundary
conditions $\rho(0)=\rho_c, \Phi(\infty)=\Lambda(\infty)=0.$

\begin{figure}
\epsfig{file=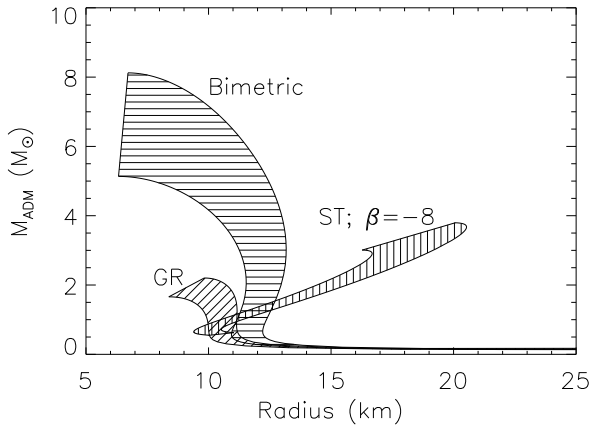,width=90mm}
\caption{Mass-radius relations for neutron stars in the GR, bimetric, 
and scalar-tensor theories. For the scalar-tensor theory, we plot the 
relation for only one value of the parameter $\beta$. The shaded regions 
show the allowed values of $M$ and $R$ for equations of state with 
stiffness between EOS~A and EOS~UU.}
\label{1}
\end{figure}

Figure \ref{1} shows the relationship between ADM mass and
neutron-star radius for the three theories under consideration. For
the case of the scalar-tensor theory, we show, as an example, the case
$\beta=-8$, which is comparable to the most negative value of $\beta$
not yet ruled out by binary pulsar timing~\cite{de96}. For each of the
three theories, we plot a shaded region corresponding to the allowed
values of mass and radius for equations of state with stiffness
between EOS~A and EOS~UU. We include only the stars stable to radial
perturbation and, for configurations in the scalar-tensor theory 
we show only the one energetically preferred.

As can be seen immediately, all three theories allow neutron stars
with masses in the currently observed range of $1.35-1.8 M_{\odot}$ to
exist \cite{tc99}. As a result, measurement of neutron star masses
alone cannot distinguish between them. This degeneracy is broken,
however, by considering the predicted radii. In the astrophysically
relevant range $M_{\text{ADM}}\geq 1.3 M_{\odot}$, the three theories
occupy mutually exclusive regions in the $(M_{\text{ADM}},R)$
space. Observations that can put limits on both the mass and radius of
a neutron star can thus constrain, without ambiguity, the permissible
set of gravity theories.

One such observation is of the surface redshift factor, $z$, relative
to infinity. This is defined by $E_{\infty}=[1/(1+z)]E_{\text{surf}}$,
where $E_{\text{surf}}$ is the energy of a photon emitted from the
stellar surface and measured at infinity with energy $E_{\infty}$. In
the GR Schwartzschild metric, this provides a direct measurement of
only the ratio $M_{\text{ADM}}/R$. In the case of the scalar-tensor
and bimetric theories the relationship between $M_{\text{ADM}}$, $R$, and
$z$ is more complicated. In all cases, however, the formula,
\begin{equation}
E_{\infty}=\left(\frac{g_{00,s}}{g_{00,\infty}}\right)^{1/2}E_{\text{surf}},
\end{equation}holds, where $g_{00,s}$ is evaluated at the surface of the star 
and $g_{00,\infty}$ is evaluated at infinity. 
\begin{figure}
\epsfig{file=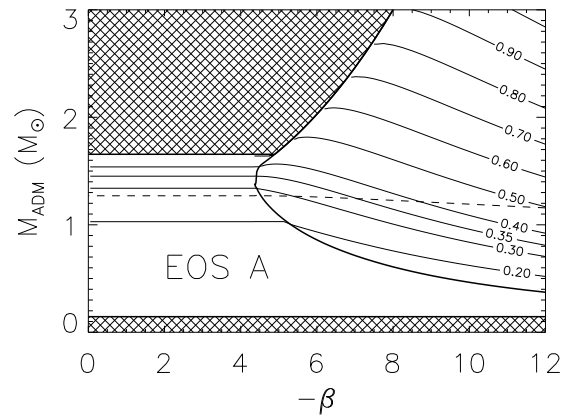,width=90mm}
\caption{Contours of constant redshift in the scalar-tensor theory, 
as a function of $M_{\text{ADM}}$ and $\beta$, for EOS~A. The dashed 
line shows the value of $M_{\text{ADM}}$, as a function of $\beta$, 
for which the baryonic mass is equal to $1.4 M_{\odot}$. The heavy 
lines on the plot show the mass limits for neutron stars and the 
bounding region where spontaneously scalarized stars are produced. 
No stable neutron stars in the crosshatched region exist.}
\label{3}
\end{figure}
\begin{figure}
\epsfig{file=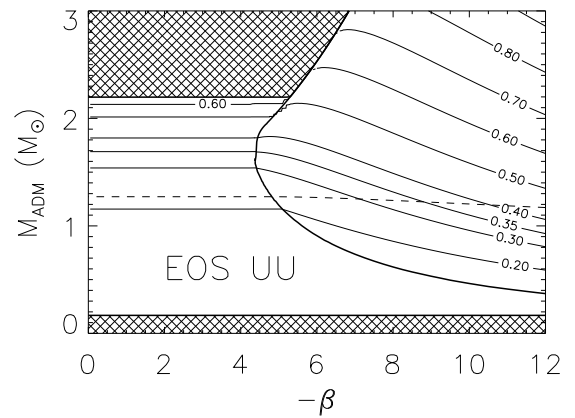,width=90mm}
\caption{Contours of constant redshift in the scalar-tensor theory, 
as a function of $M_{\text{ADM}}$ and $\beta$, for EOS~UU. See figure~\ref{3} for details.}
\label{4}
\end{figure}
\begin{figure}
\epsfig{file=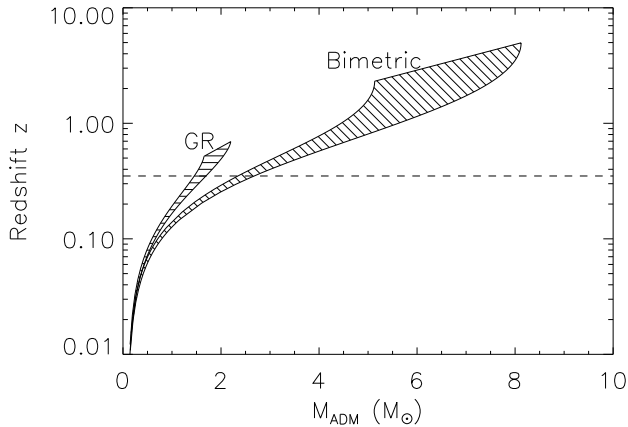,width=90mm}
\caption{Mass-redshift relations for the GR and bimetric theories. 
The dashed line shows $z=0.35$, the value reported by Cottam et al.~\cite{cpm02}.}
\label{2}
\end{figure}

Figures \ref{3} and \ref{4} show the redshift factor $z$ for the
scalar-tensor and GR theories as contours on a plot of the mass
$M_{\text{ADM}}$ versus the parameter $\beta$, for EOS~A and EOS~UU,
respectively. As discussed above, the phenomenon of spontaneous
scalarization occurs only for $\beta\lesssim -4.35$, and so above this
value the contours are parallel to the $\beta$ axis and equal to the
GR value.
The heavy lines on the plot show the upper mass limits
for neutron stars as well as the bounding region, in the
$(M_{\text{ADM}},\beta)$ space, where spontaneously scalarized stars
are produced. The dashed line shows the value of $M_{\text{ADM}}$ as a
function of $\beta$ for which the baryonic mass is equal to $1.4
M_{\odot}$. Note the $z=0.35$ contour, the value recently measured by
Cottam et al. \cite{cpm02} in the neutron star source EXO 0748--676.

The redshifts found for stars in the scalar-tensor theories differ
greatly from their general relativistic values for both equations of
state. The trend is for the redshift values to increase (nearly
always) monotonically as $\beta$ becomes more negative. If we can
place constraints on the neutron star
mass, a redshift measurement will put an upper bound on the value of
$-\beta$.

Figure \ref{2} shows the redshift factor $z$ as a function of
$M_{\text{ADM}}$ for neutron stars produced in the GR and bimetric
theories. We plot two shaded regions, one for GR and one for the
bimetric theory, that cover the range between EOS~A and EOS~UU. The
dashed line shows the recent $z=0.35$ measurement \cite{cpm02}. Even
when a GR and a bimetric star have the same redshift, the mass serves to
break the degeneracy, providing a test of strong field
gravity despite the uncertainty in the equation of state.

The recent measured redshift of $z=0.35$~\cite{cpm02} cannot by itself
constrain strong-field gravity in the theories we consider
here. However, strong limits can be placed if the mass of the neutron
star is measured directly from binary dynamics or constrained by
arguments from formation mechanisms (e.g., restricting the baryonic
mass to be above the Chandrasekhar limit for a degenerate core.) The
neutron star EXO~0748--676, for which this redshift was measured, is a
member of an eclipsing binary system with an orbital period of 3.8 h
\cite{pwg86} and hence is a prime candidate for producing a mass
measurement that can provide such a constraint. Simply requiring the
baryonic mass of the star to be greater than $1.4 M_{\odot}$ (the
Chandrasekhar limit for its degenerate progenitor) places an upper
limit on $-\beta$ of 7 (EOS~A) to 9 (EOS~UU). As expected, the
constraint on strong-field gravity depends only weakly on the equation
of state.

In addition to the redshift of atomic lines, a number of other
observational characteristics of neutron stars depend -- in
more complicated ways -- on the of nature strong-field gravity. These are
the maximum allowed spin frequency, the Eddington limit for accreting
stars, bursting behavior, cooling rates, and the
frequencies of quasi-periodic oscillations.
The constraints 
imposed by quasi-periodic oscillations will be reported in a forthcoming
paper.

The constraints on scalar-tensor discussed here have a wider
applicability than the structure of compact objects because varieties
of scalar-tensor theories naturally arise as low-energy limits of
gravitational theories in higher dimensions~\cite{ow98}. They are also
under consideration in cosmology as explanations of evidence for an
accelerating universe. To date, little attention has been given to the
implications of such theories for the properties of compact objects
that are now observable. The uncertainties in the equation of state at
ultra-high densities were thought to make such studies fruitless. Our
investigation has shown that this is not the case and that
measurements of the redshifts of surface atomic lines from neutron
stars can provide new tests of strong-field gravity.

\bigskip
We thank A. Loeb, J. Maldacena, C. Miller, and F. \"Ozel for useful
discussions. SD acknowledges the support of an NSF Graduate Research
Fellowship. DP acknowledges the support of NSF grant PHY-0070928.

\bibliographystyle{apsrev}

\begin{thebibliography}{19}
\expandafter\ifx\csname natexlab\endcsname\relax\def\natexlab#1{#1}\fi
\expandafter\ifx\csname bibnamefont\endcsname\relax
  \def\bibnamefont#1{#1}\fi
\expandafter\ifx\csname bibfnamefont\endcsname\relax
  \def\bibfnamefont#1{#1}\fi
\expandafter\ifx\csname citenamefont\endcsname\relax
  \def\citenamefont#1{#1}\fi
\expandafter\ifx\csname url\endcsname\relax
  \def\url#1{\texttt{#1}}\fi
\expandafter\ifx\csname urlprefix\endcsname\relax\def\urlprefix{URL }\fi
\providecommand{\bibinfo}[2]{#2}
\providecommand{\eprint}[2][]{\url{#2}}

\bibitem[{\citenamefont{Cottam et~al.}(2002)\citenamefont{Cottam, Paerels, and
  Mendez}}]{cpm02}
\bibinfo{author}{\bibfnamefont{J.}~\bibnamefont{Cottam}},
  \bibinfo{author}{\bibfnamefont{F.}~\bibnamefont{Paerels}}, \bibnamefont{and}
  \bibinfo{author}{\bibfnamefont{M.}~\bibnamefont{Mendez}},
  \bibinfo{journal}{Nature} \textbf{\bibinfo{volume}{420}}, \bibinfo{pages}{51}
  (\bibinfo{year}{2002}).

\bibitem[{\citenamefont{Sanwal et~al.}(2002)\citenamefont{Sanwal, Pavlov,
  Zavlin, and Teter}}]{spzt02}
\bibinfo{author}{\bibfnamefont{D.}~\bibnamefont{Sanwal}},
  \bibinfo{author}{\bibfnamefont{G.~G.} \bibnamefont{Pavlov}},
  \bibinfo{author}{\bibfnamefont{V.~E.} \bibnamefont{Zavlin}},
  \bibnamefont{and} \bibinfo{author}{\bibfnamefont{M.~A.} \bibnamefont{Teter}},
  \bibinfo{journal}{\apj} \textbf{\bibinfo{volume}{574}}, \bibinfo{pages}{L61}
  (\bibinfo{year}{2002}).

\bibitem[{\citenamefont{Damour and Esposito-Far\`{e}se}(1992)}]{de92}
\bibinfo{author}{\bibfnamefont{T.}~\bibnamefont{Damour}} \bibnamefont{and}
  \bibinfo{author}{\bibfnamefont{G.}~\bibnamefont{Esposito-Far\`{e}se}},
  \bibinfo{journal}{Class. Quantum Grav.} \textbf{\bibinfo{volume}{9}},
  \bibinfo{pages}{2093} (\bibinfo{year}{1992}).

\bibitem[{\citenamefont{Rosen}(1973)}]{r73}
\bibinfo{author}{\bibfnamefont{N.}~\bibnamefont{Rosen}}, \bibinfo{journal}{J.
  Gen. Rel. and Grav.} \textbf{\bibinfo{volume}{4}}, \bibinfo{pages}{435}
  (\bibinfo{year}{1973}).

\bibitem[{\citenamefont{Will and Eardley}(1977)}]{we77}
\bibinfo{author}{\bibfnamefont{C.~M.} \bibnamefont{Will}} \bibnamefont{and}
  \bibinfo{author}{\bibfnamefont{D.~M.} \bibnamefont{Eardley}},
  \bibinfo{journal}{\apj} \textbf{\bibinfo{volume}{212}}, \bibinfo{pages}{L91}
  (\bibinfo{year}{1977}).

\bibitem[{\citenamefont{Damour and Esposito-Far\`{e}se}(1993)}]{de93}
\bibinfo{author}{\bibfnamefont{T.}~\bibnamefont{Damour}} \bibnamefont{and}
  \bibinfo{author}{\bibfnamefont{G.}~\bibnamefont{Esposito-Far\`{e}se}},
  \bibinfo{journal}{\prl} \textbf{\bibinfo{volume}{70}}, \bibinfo{pages}{2220}
  (\bibinfo{year}{1993}).

\bibitem[{\citenamefont{Will}(1993)}]{w93}
\bibinfo{author}{\bibfnamefont{C.}~\bibnamefont{Will}},
  \emph{\bibinfo{title}{Theory and Experiment in Gravitational Physics}}
  (\bibinfo{publisher}{Cambridge University Press}, \bibinfo{year}{1993}).

\bibitem[{\citenamefont{Harada}(1998)}]{h98}
\bibinfo{author}{\bibfnamefont{T.}~\bibnamefont{Harada}},
  \bibinfo{journal}{\prd} \textbf{\bibinfo{volume}{57}}, \bibinfo{pages}{4802}
  (\bibinfo{year}{1998}).

\bibitem[{\citenamefont{{Cook} et~al.}(1994)\citenamefont{{Cook}, {Shapiro},
  and {Teukolsky}}}]{cst94}
\bibinfo{author}{\bibfnamefont{G.~B.} \bibnamefont{{Cook}}},
  \bibinfo{author}{\bibfnamefont{S.~L.} \bibnamefont{{Shapiro}}},
  \bibnamefont{and} \bibinfo{author}{\bibfnamefont{S.~A.}
  \bibnamefont{{Teukolsky}}}, \bibinfo{journal}{\apj}
  \textbf{\bibinfo{volume}{424}}, \bibinfo{pages}{823} (\bibinfo{year}{1994}).

\bibitem[{\citenamefont{Pandharipande}(1971)}]{eosa}
\bibinfo{author}{\bibfnamefont{V.~R.} \bibnamefont{Pandharipande}},
  \bibinfo{journal}{Nucl. Phys. A} \textbf{\bibinfo{volume}{174}},
  \bibinfo{pages}{641} (\bibinfo{year}{1971}).

\bibitem[{\citenamefont{Wiringa et~al.}(1988)\citenamefont{Wiringa, Fiks, and
  Fabrocini}}]{eosuu}
\bibinfo{author}{\bibfnamefont{R.~B.} \bibnamefont{Wiringa}},
  \bibinfo{author}{\bibfnamefont{V.}~\bibnamefont{Fiks}}, \bibnamefont{and}
  \bibinfo{author}{\bibfnamefont{A.}~\bibnamefont{Fabrocini}},
  \bibinfo{journal}{\prc} \textbf{\bibinfo{volume}{38}}, \bibinfo{pages}{1010}
  (\bibinfo{year}{1988}).


\bibitem[{\citenamefont{{Damour} and {Esposito-Far{\` e}se}}(1996)}]{de96}
\bibinfo{author}{\bibfnamefont{T.}~\bibnamefont{{Damour}}} \bibnamefont{and}
  \bibinfo{author}{\bibfnamefont{G.}~\bibnamefont{{Esposito-Far{\` e}se}}},
  \bibinfo{journal}{\prd} \textbf{\bibinfo{volume}{54}}, \bibinfo{pages}{1474}
  (\bibinfo{year}{1996}).

\bibitem[{\citenamefont{{Thorsett} and {Chakrabarty}}(1999)}]{tc99}
\bibinfo{author}{\bibfnamefont{S.~E.} \bibnamefont{{Thorsett}}}
  \bibnamefont{and}
  \bibinfo{author}{\bibfnamefont{D.}~\bibnamefont{{Chakrabarty}}},
  \bibinfo{journal}{\apj} \textbf{\bibinfo{volume}{512}}, \bibinfo{pages}{288}
  (\bibinfo{year}{1999}).

\bibitem[{\citenamefont{{Barziv} et~al.}(2001)\citenamefont{{Barziv}, {Kaper},
  {Van Kerkwijk}, {Telting}, and {Van Paradijs}}}]{b01}
\bibinfo{author}{\bibfnamefont{O.}~\bibnamefont{{Barziv}}},
  \bibinfo{author}{\bibfnamefont{L.}~\bibnamefont{{Kaper}}},
  \bibinfo{author}{\bibfnamefont{M.~H.} \bibnamefont{{Van Kerkwijk}}},
  \bibinfo{author}{\bibfnamefont{J.~H.} \bibnamefont{{Telting}}},
  \bibnamefont{and} \bibinfo{author}{\bibfnamefont{J.}~\bibnamefont{{Van
  Paradijs}}}, \bibinfo{journal}{\aa} \textbf{\bibinfo{volume}{377}},
  \bibinfo{pages}{925} (\bibinfo{year}{2001}).

\bibitem[{\citenamefont{{Orosz} and {Kuulkers}}(1999)}]{ok99}
\bibinfo{author}{\bibfnamefont{J.~A.} \bibnamefont{{Orosz}}} \bibnamefont{and} \bibinfo{author}{\bibfnamefont{E.}~\bibnamefont{{Kuulkers}}},
  \bibinfo{journal}{MNRAS} \textbf{\bibinfo{volume}{305}}, \bibinfo{pages}{132}
  (\bibinfo{year}{1999}).

\bibitem[{\citenamefont{{Parmar} et~al.}(1986)\citenamefont{{Parmar}, {White},
  {Giommi}, and {Gottwald}}}]{pwg86}
\bibinfo{author}{\bibfnamefont{A.~N.} \bibnamefont{{Parmar}}},
  \bibinfo{author}{\bibfnamefont{N.~E.} \bibnamefont{{White}}},
  \bibinfo{author}{\bibfnamefont{P.}~\bibnamefont{{Giommi}}}, \bibnamefont{and}
  \bibinfo{author}{\bibfnamefont{M.}~\bibnamefont{{Gottwald}}},
  \bibinfo{journal}{\apj} \textbf{\bibinfo{volume}{308}}, \bibinfo{pages}{199}
  (\bibinfo{year}{1986}).


\bibitem[{\citenamefont{{Overduin} and {Wesson}}(1997)}]{ow98}
\bibinfo{author}{\bibfnamefont{J.~M.} \bibnamefont{{Overduin}}}
  \bibnamefont{and} \bibinfo{author}{\bibfnamefont{P.~S.}
  \bibnamefont{{Wesson}}}, \bibinfo{journal}{Phys. Rep.}
  \textbf{\bibinfo{volume}{283}}, \bibinfo{pages}{303} (\bibinfo{year}{1997}).

\end{thebibliography}

\end{document}